\documentstyle[prl,aps,multicol,epsfig,tabularx]{revtex} 

\begin{document} 

\draft
\title{Swarming ring patterns in bacterial colonies exposed to ultraviolet radiation}
\author{Anna M. Delprato$^{1,2}$, Azadeh Samadani$^{1}$,
A. Kudrolli$^{1,}$\cite{Corr},  and L.S. Tsimring $^{3}$}

\address{$^{1}$ Department of Physics, Clark University, Worcester, MA 
01610\\
$^{2}$ Department of Biology, Clark University, Worcester, MA 01610\\
$^{3}$ Institute for Nonlinear Science, University of California, San Diego, CA
92093-0402} 

\date{\today}  

\maketitle

\begin{abstract}
We report a novel morphological transition in a {\it Bacillus subtilis}
colony initially growing under ambient conditions, after ultraviolet
radiation exposure. The  bacteria in the central regions of the colonies
are observed to migrate towards the colony edge forming a ring during
uniform spatial exposure. When the radiation is switched off, the
colonies were observed to grow both inward into the evacuated regions as
well as outward indicating that the pattern is not formed due to
depletion of nutrients at the center of the colony. We also propose a
reaction-diffusion model in which waste-limited chemotaxis initiated by 
the UV radiation leads to the observed phenomenology.
\end{abstract}



\begin{multicols}{2} 
Bacterial colony growth on semi-solid agar has served as a model system
to understand self-organization of microscopic biological elements to
form complex patterns~\cite{matsushita90,budrene91,ben-jacob94}.
Although interaction between individuals may be well known, it is
difficult to determine how complex patterns form for example due to
chemotaxis. Random walk and reaction diffusion models have been also
developed to find the minimal mechanisms that lead to the formation of
the observed
patterns~\cite{ben-jacob94,tsimring95,brenner98,kozlovsky99,tyson98}. In
most of these studies, changes in the environment occur due to the slow
depletion of nutrients and excretion of waste. 

The modeling of the response of living organisms to a change in
environment is an important issue of current interest. An example is the
effect of ultraviolet (UV) radiation on biological systems prompted by
stratospheric ozone depletion. Recent studies have modeled the response
to externally imposed change on bacterial colonies using the Fisher wave
equation with added convection terms~\cite{shnerb98,dahmen00}.
Experiments using ultraviolet illumination pattern changes on {\it
Bacillus subtilis} colonies have shown the relevance of such
modeling~\cite{neicu00}. However, a delay in the response of the
bacterial colony to changes in the environment was also noted. These
studies raise interesting questions on how to model biological systems
that may undergo growth mode changes and show history dependence in
response to environmental changes. 

In this paper, we report the spatio-temporal response of colonies of
{\it Bacillus Subtilis} growing on nutrient rich agar to a sudden change
in the environment. {\it Bacillus Subtilis} is a 9$\mu$m $\times$
2$\mu$m motile bacterium that can be often found in soil. The change is
hostile and is implemented by exposing the colony to uniform UV
radiation. The colony which grows uniformly before exposure to UV
radiation, is not observed to simply stop due to the radiation but
undergo a morphology change. Most of the bacteria are observed to
migrate to the edge of the colony even though this does not help the
bacteria to evade the radiation. To systematically study the 
spatio-temporal response of the colony, we
measured the change in the bacterial density as a function of the
strength of the UV radiation for the wild type and DNA repair mutants,
uvrA and uvrA recA of the bacterium. We also propose a simple
reaction-diffusion model which reproduces most of the observed
phenomenology. The model operates with three densities: bacteria, waste,
and chemoattractant, and describes a waste-limited chemotaxis initiated
by the UV radiation. The bacteria migrate towards the edges because the waste concentration is reduced at the edges of the colony. 

Experiments were performed in 15 cm diameter Plexiglas petri dishes
containing a thin layer of nutrient agar (7 grams/liter of bacto peptone
and 3 grams/liter agar). Under these conditions the growth of the
bacterial colony can be approximately modeled by the Fisher wave
equation~\cite{wakita94,neicu00}. The bacterial colony was initiated
with an inoculating needle at a single point source at the center of the
petri dish and the bacteria grew in a thin layer at the surface of the
medium. The colony is first allowed
to grow in the absence of UV radiation. The initial adaptation time and
subsequent growth rate of the colony depends on the temperature and for
most of the experiments a temperature of 24$^o$C was used. At this
temperature, the colony front was observed to grow slowly for the first
26 hours and then increased reaching a constant front speed $v_f = 0.3
\mu$m/s. The colony was then exposed to uniform UV radiation over a time
interval $t_{uv}$ using a  fluorescent bulb, (Philips F15T8) which
radiates in the range of 310-400 nm (UV-A and UV-B).  An example of the spatio-temporal response in a colony of {\it Bacillus
subtilis}, uvrA  under UV radiation is shown in Fig.~\ref{fig1}(a)-(c).

We note that the circular symmetry of the patterns is a consequence of
the point inoculation and the subsequent disk shaped colony which
develops before the colony is exposed to UV radiation. If the bacteria
is grown in a ribbon shape using a line inoculation pattern, they still migrate to the edges.
For simplicity, we performed quantitative experiments with colonies using point
inoculation.

To further investigate the cause of the migration of the bacterial
population, we observed the recovery of the colony after exposure to
radiation. Images of the recovering colony are shown in
Fig.~\ref{fig1}(d)-(f) and it can be noted that the colony recovers almost
uniformly in both directions. The colony grows
outward to fresh regions and also inward into the evacuated central
regions. 

The radius of the circular colony before exposure, during exposure and after exposure was plotted as a function of time in Fig.~\ref{fig1-front}. We
observed that the growth of the front of the colony during exposure is negligible
compared to the growth before exposure. After radiation is switched off, it can be
observed that cell division resumes following  a ``recovery" time $t_r$, which
increases with the strength of the radiation. Although, the front speed on the outside edge is greater than
the inside edge, clearly the central
regions of the colony are hospitable. Therefore, the initial evacuation
of the colony at the center is not due to depletion of nutrients. Thus
the formation of the swarm ring like patterns has clearly a different
mechanism than observed previously~\cite{budrene95,tyson98}.

To quantitatively show the rearrangement of the bacterial colony as a
function of time, the bacterial density $n(r)$ as a function of distance
$r$ from the center is plotted in Fig.~\ref{uv-strength}(a). The density
of the bacteria scales linearly with light intensity and therefore the
density distribution of the colony is obtained from the images using a
CCD camera. At approximately 24 hours the density of the bacteria is
considerably higher at the edge of the colony compared to the colony
center. After 70 hours the bacteria almost completely evacuate the
center of the colony. 

In Fig.~\ref{uv-strength}(b), we show the dependence of the strength of the UV
radiation $I$ on the colony. The cross section of the colony was plotted
after the colonies were allowed to grow and were exposed to UV radiation
for the same time duration $t_{uv}$. The front of the colony was also
observed to advance faster with respect to the higher intensity, but was
still much slower compared to the $v_f$ before exposure to UV light (see
Table 1). The total population of the colony as a function of time 
was measured by integrating the population
densities and is shown in Fig.~\ref{uv-strength}(c). For lower $I$, the population when exposed to radiation increases at a small rate initially, and then decreases. However, for higher $I$ the total population always decreases. 

Qualitatively similar phenomena was observed in the two other strains of
{\it Bacillus subtilis}. Although the density was observed to increase in the wild type strain
near the edge of the colony, the center of the colony was not observed
to vacate completely. At the same UV radiation strength, the total
population and the front were also observed to grow at a faster rate
compared to uvrA. In the case of uvrA recA, the colony was observed to
have a higher density at the edges but the total population decreased
strongly with time. Thus the behavior of the recA strain can be thought of as intermediate between that observed in the wild and the uvrA recA strains. This is consistent with the fact that 
while all mechanisms of DNA repair are
present in the wild type, one mechanism of repair is absent in uvrA, and
two are absent in the urvA recA strain~\cite{jagger85,munakata89,hanlin85,lewin97}. 
Therefore the rate and strength of the density
distribution appears to be related to the efficiency of the DNA repair mechanism.

In an attempt to elucidate the mechanism of this morphological
transition, we propose the following simple model\cite{model}. The model
consists of three reaction-diffusion equations for bacteria concentration 
$b({\bf r},t)$, waste concentration $w({\bf r},t)$, and chemoattractant concentration $c({\bf r}, t)$: 
\begin{eqnarray}
\partial_t b &=& f b(1-w) -\mu (1-f)b - \nabla\cdot{\bf J}_c+D_b\nabla^2 b 
\label{b}\\
\partial_t w &=& f b(1-w) +D_w\nabla^2 w
\label{w}\\
\partial_t c &=& (1-f)b(1-w) +D_c\nabla^2 c
\label{c}
\end{eqnarray}

The first terms in the r.h.s. of Eqs.(\ref{b}),(\ref{w}) describe the
bacterial growth and the accompanying waste production, respectively.
The bacterial growth is limited by the local concentration of waste $w$
and saturates when waste concentration approaches 1.  Constant $f=1$ in
the absence of the UV radiation and becomes small when UV radiation is
switched on.  According to Eq.(\ref{c}), when $f<1$, bacteria emit
chemoattractant $c$ at the rate which is proportional to the bacteria
concentration and also is limited by the waste concentration. The term
$-\mu(1-f)b$ in Eq.(\ref{b}) is responsible for the slow destruction of
bacteria by the UV radiation observed experimentally.  The chemotactic
flux ${\bf J}_c=\alpha b(1-b)\nabla c$ is directed towards the gradient
of the concentration of chemical $c$, and it saturates at large
bacterial concentration $b\to 1$ (hard core repulsion).  The
last terms in Eqs.(\ref{b})-(\ref{c}) describe linear diffusion of the
components.  For simplicity, we assume that the diffusion constants of
all three fields are equal and rescale them to unity.  

The initial colony expansion without UV ($f=1$) is
described by the first two equations with ${\bf J}_c=0$. Asymptotically
$w-b\to 0$, and the system is reduced to a single
Kolmogorov-Petrovsky-Piskunov-Fisher equation \cite{KPPF} 
\begin{equation}
\partial_t b = f b(1-b)+\nabla^2 b,
\label{kppf}
\end{equation}
which has a propagating front solution with the front speed $v_f=2$. 

The simplest way to model the effect of UV radiation is to set $f=0$ at
some time $t=t_0$.  The growth of the bacteria is arrested, and the
colony expansion stops. The subsequent dynamics of the population is
determined by the interplay of diffusion and chemotaxis.  In the model,
the bacteria immediately begin to emit chemoattractant $c$ in the regions
where waste $w$ is below the limiting level $w=1$, i.e. near the edges
of the colony. A ring of $c$ is formed, and the exponentially decaying
tails of this ring extend towards the center of the colony, thereby
provoking the bacteria to concentrate near the edges of the colony. 

We integrated the model (\ref{b})-(\ref{c}) numerically in a square
150x150 with periodic boundary conditions starting from a small
radially-symmetric non-zero perturbation of $b$ in the center for
$\mu=0.02, \alpha=2$. Variations in the strength of the UV radiation
can be simulated by using different small values of $f(t>t_0)$ instead
of zero.  In a typical simulation we switched $f$ from 1 to 0.1 at
$t=20$, when the radius of the colony was $r\approx 21$.  The resulting
ring pattern and the radial density distribution as a function of time
are shown in Fig.~\ref{model}.  As in the experiment, after the UV radiation
is turned on, the population starts to assemble near the edge (cf.
Fig.~\ref{uv-strength}(a)). Due to slow bacterial growth, the
outer front continues to expand, however, at a much slower rate. The total
population size slightly grows initially at $t>t_0$, but later declines.
All this agrees completely with observed phenomenology.  A more detailed
discussion of the model and its comparison with experiments will be
reported elsewhere.

In summary, colonies of {\it Bacillus subtilis} are observed to show a
swarming ring like pattern when exposed to UV radiation. Thus the mode
in which the bacterial colonies grow after exposure to UV radiation is
very different from the Fisher wave like mode in which it grows prior to
exposure. Using reaction-diffusion modeling, we have shown that the migration 
of colony under UV radiation is related to chemotaxis which is limited by  
the waste that {\it Bacillus subtilis} produces. 

We thank Nobou Munakata of the National Cancer Center Research Institute
Tokyo, Japan for supplying the mutant strains of {\it Bacillus
subtilis}. We thank T. Neicu and J. Newburg-Rinn for help with the
experiments. L. T. acknowledges partial support from 
the U.S. DOE under grant DE-FG03-95ER14516 and  DE-FG03-96ER14592.

\begin{table}
\begin{center} 
\begin{tabular}{c c c c} 
\hline  
$I$ \,\,  & $ v_f$ (initial)  & $ v_f $ (UV) &  $ v_g $ (UV)\\ 

W/m$^2$ \,\,  & $\mu$m s$^{-1}$\,\, & $\mu$m s$^{-1}$ & $\mu$m s$^{-1}$ \\

\hline \hline 

30  & 0.3 & 0.004  &  0.016-0.020 \\

12  & 0.3 & 0.006  & 0.013-0.016 \\ 

 7  & 0.3 & 0.010  & 0.011-0.013 \\

\hline 

\end{tabular}
\end{center}
\caption{The front and group velocity of {\it Bacillus subtilis} uvrA as a function of intensity $I$ of the UV radiation.}
\end{table}

\begin{figure}
\begin{center}  
\end{center}
\caption{(a)-(c) The images of {\it Bacillus subtilis}, uvrA as a function of time $t_{uv}$ under ultraviolet (UV) radiation. (a) $t_{uv}$ = 0 h (b) $t_{uv}$ = 25 h and (c) $t_{uv}$ = 45 h. The UV radiation of strength $I = 30 {\rm W/m^2} $ is switched on after the colony was allowed to grow for time $t = 50$ hours. The bacterial population is observed to migrate to the edge of the colony. The bright point at the center of the colony corresponds to the point of inoculation. (d)-(f) Recovery images of the bacterial population as a function of time $t_{r}$ after the UV radiation is switched off. The bacterial colony was allowed to recover for (d) $t_{r}$ = 30 h (e) $t_{r}$ = 40 h and (f) $t_{r}$ = 55 h. The bacterial population is observed to grow both outward as well as into the vacated inner region.}
\label{fig1} 
\end{figure}


\begin{figure}
\begin{center} 
\centerline{\epsfig{file=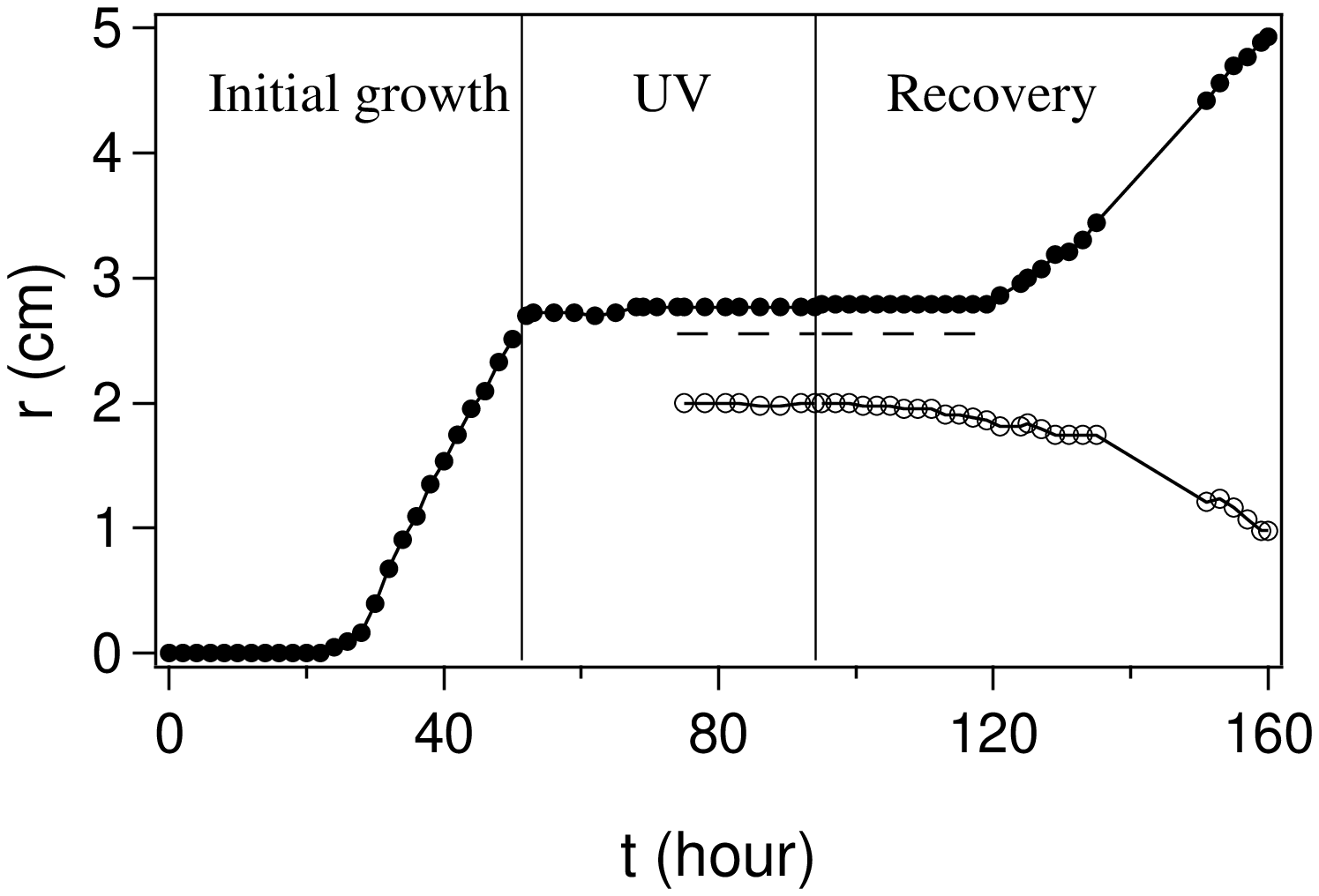,width=6 cm}}
\end{center}
\caption{The radius of the colony as a function of time from the inoculation 
($I = 30 {\rm W/m^2} $). The outer edge of the colony is plotted using ($\bullet$), and the inner edge using ($\circ$). The bacterial colony is observed to recover both inwards and outwards in the absence of UV radiation. The dashed line corresponds to the point where population density has decreased to less than 10 \% of the maximum value at the edge.}
\label{fig1-front} 
\end{figure} 


\begin{figure}
\begin{center}
\centerline{\epsfig{file=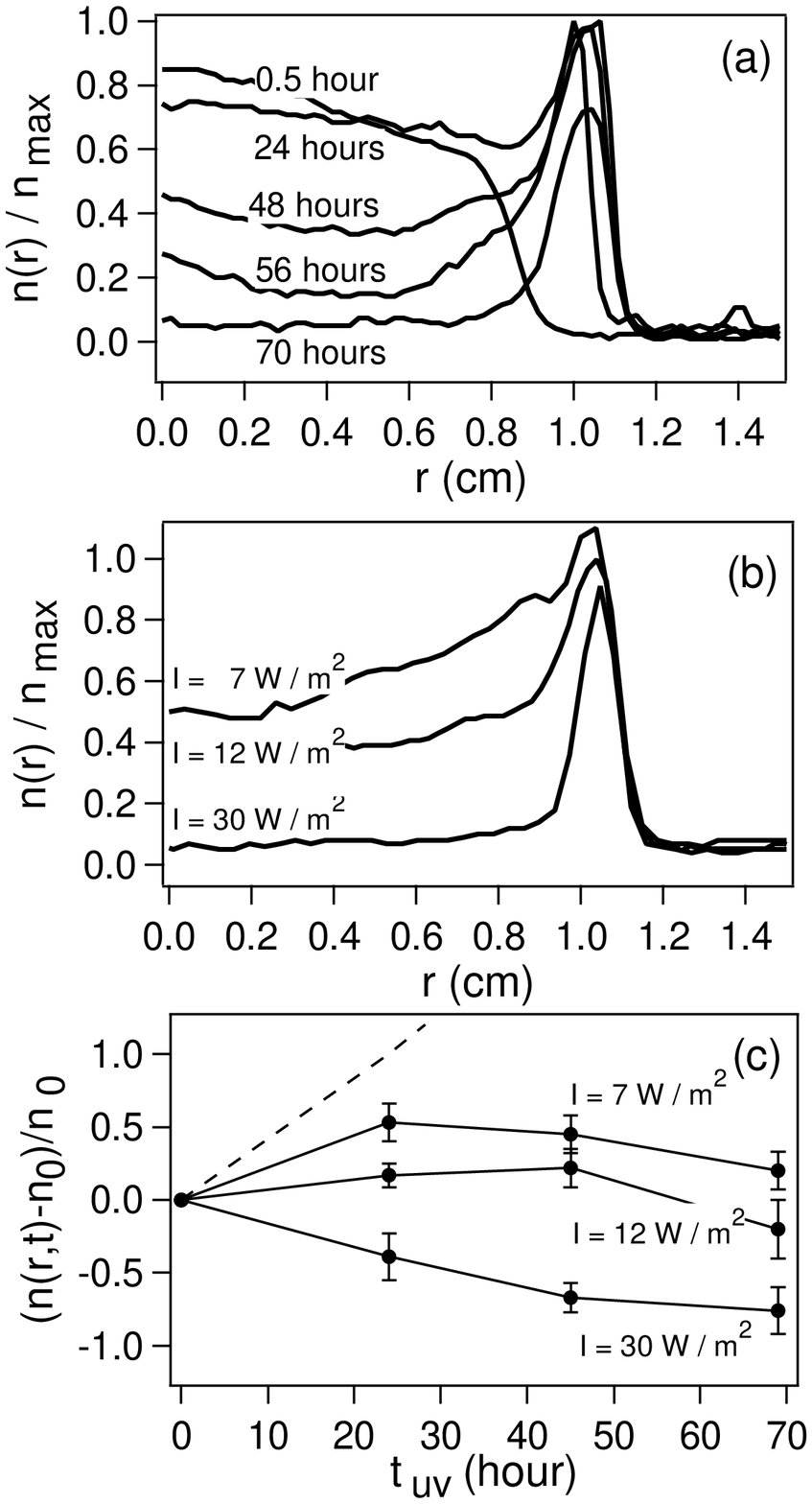, width=6 cm}} 
\end{center}
\caption{(a) The population density $n(r)/n_{max}$ as a function of distance $r$ from the inoculation point and as a function of time for $I = 12 {\rm W/m^2} $. $n_{max}$ is the maximum density at $t_{uv} =0$. (The data is shifted by approximately 2 mm from the inoculation point to avoid the puncture mark.)
(b) The $n(r)/n_{max}$ as a function of radiation strength $I$ at for 46 hours. The ring is observed to form faster as the intensity of UV radiation is increased. The colony front for the three cases have been offset to align at the colony edge for clarity. (c) The fractional change of the total population as a function of time after the UV radiation is turned on for the three intensities. The dashed line is the fractional increase in the population in the absence of UV radiation.}
\label{uv-strength} 
\end{figure} 


\begin{figure}
\begin{center} 
\end{center}
\caption{The bacterial density $b(r)$ as a function of distance $r$ from the center at 10 unit time intervals after the radiation was effectively turned on in Eqs.(\ref{b})-(\ref{c}) for $\mu=0.02$, $\alpha=2$. Inset: A ring formed using the proposed model is similar to that observed in the experiments.}
\label{model} 
\end{figure} 

\end{multicols}


\begin{thebibliography}{}
\bibitem{model} The model which we propose here, is similar to the model
for the aggregation bacterial patterns reported in
Ref.\protect\cite{tsimring95}.  The important difference
however is that in Ref.~\cite{tsimring95},  the chemoattractant emission was
triggered by the waste production, whereas in the present situation, the
chemoattractant emission is triggered by the UV radiation and is limited
by waste. This factor prevents the chemotactic instability that would lead
to bacteria clumping in multiple rings or spots.
\end{thebibliography}
\end{document}